\documentclass{icrc}

\usepackage{times}
\usepackage{graphicx} 

\begin{document}

\title{Status of Project GRAND's Proportional Wire Chamber Array}
\author[1]{J. Poirier}
\author[1]{D. Baker}
\author[1]{J. Barchie} 
\author[1]{C. D'Andrea}
\author[1]{M. Dunford}
\author[1]{M. Green}
\author[1]{J. Gress}
\author[1]{T. Lin}
\author[1]{D. Race}
\author[1]{R. Skibba}
\author[1]{G. VanLaecke}
\author[1]{M. Wysocki}

\affil[1]{Center For Astrophysics at Notre Dame, Physics Dept., University 
of Notre Dame, Notre Dame, Indiana 46556 USA}

\correspondence{John Poirier (Poirier@nd.edu)}

\runninghead{Poirier: Asymmetries}
\firstpage{1}
\pubyear{2001}

\maketitle

\begin{abstract}
Project GRAND is an extensive air shower array of proportional wire chambers.
It has 64 stations in a 100~m x 100~m area; each station has eight planes of
proportional wire chambers with a 50 mm steel absorber plate above the
bottom two planes.  This arrangement of  
planes, each 1.25 square meters of area,
allow an angular measurement for each track to 0.25$^\circ$ in each of two
 projections.  The steel absorber plate allows a measurement of
the identity of each muon track to 96\% accuracy.  Two data-taking
triggers allow data to be simultaneously taken for a) extensive air showers
(multiple coincidence 
station hits) at about 1 Hz and b) single muons (single tracks of
identified muons) at 2000 Hz.  Eight on-line computers pre-analyze the
single track data and store the results on magnetic tape in compacted form
with a minimum of computer dead-time.  One additional computer
reads data from the shower triggers and records this raw data on a
separate magnetic tape with no pre-analysis.
\end{abstract}

\section{Introduction} 

Project GRAND utilizes a rather different technique of studying 
cosmic ray showers.  Instead of the traditional method of 
measuring the orientation of the shower front by means of timing counters, 
GRAND uses eight planes of proportional wire chambers stacked 
vertically on top of each other with a steel plate above the bottom two 
planes to geometrically measure the angles of the secondary tracks in the 
shower.  The steel absorber plate allows muon tracks to be identified.    

By locating four points (three in the case of electron tracks 
which do not make it through the steel plate)  
on a track trajectory 
in the x plane and four in the y plane, the angle of the secondary tracks 
are measured in these two orthogonal planes yielding its three-dimensional 
orientation in space.  The angle of the primary cosmic 
ray is obtained by averaging the angles of the many secondary tracks of 
the associated extensive air shower.  In averaging, the 
accuracy increases for showers with more secondary tracks.    
 
\section{Experimental Array} 

Each of the 64 stations  
is composed of four pairs of orthogonal planes.  Each plane contains 80 
detection wires; the planes are aligned to within $\pm0.2^\circ$ 
with the north/south or east/west directions.  
Adjacent planes are constructed orthogonal to one another to 
$\pm0.1^\circ$.  
Originally the array was designed to study extensive 
air showers produced by cosmic ray primaries $\geq$~100 TeV.  However,  
since each secondary track's angle is measured and its identity determined,
GRAND obtains large quantities of data on single muon 
tracks along with the extensive air shower data.  
The experiment now runs two simultaneous triggers: 
a) a coincidence trigger of $\ge$3 stations for data on 
extensive air showers, and 
b) a single-track trigger which collects all 
tracks stored in all stations 280 times a second; 
these tracks are about 75\% muon tracks.  
The former trigger is about $\sim$1~Hz and the latter $\sim$~2000 Hz.  
These data are stored independently on two different magnetic tape 
drives.  

\subsection{Muon Identification} 

There is a 50 mm steel plate above the bottom pair of planes; electrons scatter, 
stop, or shower because of the steel absorber plate but the higher mass 
muons are relatively unaffected.  
Because an electron 
will be misidentified as a muon approximately 4\% of the time and a muon 
misidentified as an electron also 4\% of the time, this arrangement 
allows Project GRAND to differentiate muon tracks with 96\% precision while 
retaining 96\% of them.  
Given the 80 channels of proportional wires in a plane and the vertical 
separation between the planes, GRAND is able to 
measure the direction of a muon track to 0.26$^\circ$, on average, in each 
of two orthogonal planes.  This geometrical arrangement of planes 
has a projected angle sensitivity cutoff-angle of $63^\circ$ from vertical.  
The muon threshold energy is 0.1~GeV for vertical 
tracks, increasing as 1/cos$\phi$ for $\phi$ inclined from vertical.    

\subsection{Electronics} 

The electronics was constructed mostly of CMOS integrated circuit chips  
made possible because of the low instantaneous rate of cosmic 
rays.   The use of shift register memory with parallel inputs 
and serial outputs allowed the data to be sent to the central 
electronics trailer on a single data cable for an entire hut with 
its 640 bits (eight planes) of data information.  
The use of CMOS chips allows for low power consumption, a large 
amount of logic in a single chip, and low cost.

The 640 bits of information from the 
eight planes of a station are read serially down a single data cable at a rate 
of 12 MHz.
The data from all 64 stations are read in parallel in 70 microsec 
into the central data 
acquisition area (a trailer obtained from government surplus). 
The master computer looks for a new event.  When it arrives, it 
determines if there are $\ge$3 huts in coincidence; if so, the master 
CPU reads the event into its memory and thence  
to the shower magnetic tape.  
If this criteria is not meet , it assigns 
a slave CPU to read in the data, 
analyze it for single tracks in a station, 
and store any muon tracks it finds in its internal buffer memory.  
The master 
CPU determines if any slave CPU's buffer memory is full and, if so, writes  
its entire 900 muon 
buffer to the single-muon magnetic tape.

As stated above, 
the single muon data is preanalyzed.  This analysis examines each station 
for a single hit in all eight planes (allowing for two adjacent hits 
which, for a track passing near the middle between two cells, happens about 
10\% of the time).  
Upon finding eight single hits in each of the 8 planes of a hut, 
it stores the location of 
each of the wire-hits in buffer memory.  When the memory reaches 900 
muons, it is then written in a single file to magnetic tape (95\% of these 
data fit a straight track on offline analysis).  
Eight on-line computer nodes working in sequence 
minimize the deadtime associated with sorting through the total of 
40960 bits of information from the entire field for each event read.  

The data is stored on 
two drives, one for shower data and one for single muon 
data.  The shower data is written directly to tape with no pre-analysis; 
it is rather sparse in hits and, when written in compression mode, stores 
several weeks of data on a single tape. 

The output of single muon data from all eight CPU-s are written to another 
tape drive.  This drive has a backup tapedrive arranged such that, 
when full, the computer automatically switches 
to the second drive (and vice versa).  Since each drive holds 2.3 day's 
data, it is only necessary to replace a tape every two or three 
days in order to keep the experiment running continuously.

\subsection{Gas} 
The PWCs require a slow flow rate of 80\% argon plus 20\% carbon dioxide gas 
mixture to maintain the purity of the gas in the PWCs.  
The flow is such that one T-cylinder of gas furnishes gas 
for 512 planes of PWCs for two days.  The gas system is so configured 
that when one tank is emptied, a second backup tank is automatically 
switched unto the sytem; thus the sequence of two cylinders lasts for 
four days.
  
\subsection{Acceptance} 

Since Project GRAND is not sensitive to the whole sky (its cutoff 
projected 
angle is 63$^\circ$ from zenith); like other ground-based detectors, 
it is more sensitive to cosmic rays coming from near its zenith. 
GRAND's acceptance, $Accept$, depends on the track angle given by:   
\begin{equation}
Accept=[1-0.537\tan\phi_x][1-0.537\tan\phi_y]\cos^3\phi
\end{equation}
where the angle $\phi$ is the muon's angle from 
the vertical or zenith direction, $\phi_x$ is the projection of $\phi$ 
upon the xz-plane and $\phi_y$ is the yz-projection. 
It combines two geometrical 
factors, 
a $cos \phi$ factor for the projection of the muons 
unto the zenith direction, and cos$^{2}\phi$ describes   
muon absorption 
in the atmosphere due to the increased path length in air for 
muons inclined from the vertical.   
The geometrical factor in equation one arises from the 
arrangement of several horizontal proportional wire planes 
placed above each other together with the demand that a track traverse 
both the top and bottom planes.  
Each PWC plane has 1.25 m$^2$ of active area.  
\section{The Proportional Wire Chambers} 
Mass production of the proportional wire chambers allows them to be built 
with high precision, uniformity of characteristics, and low cost per unit.  
In studying 
cosmic rays it is important to have a large detector area 
 (consistent with available resources); thus cost is important.  
The original goal was to build these detectors at 0.1 the cost of 
prior high energy physics experiments.  This goal 
was met in both the chamber and in the associated electronics.  
Examples of cost savings in the proportional wire chamber construction are  
the use of glass instead of a epoxy-glass composite 
and the elimination of sockets for the attachment of 
the electronics boards.

There are eight planes of proportional wire chambers per station (hut) 
arranged vertically above each other with 100 mm separation between the 
planes of a given projection.  They are arranged with four in the x-plane 
(wires running NS) and four in the y-plane (wires running EW).  A 50 mm 
steel plate is located above the bottom two planes.  There are 80 cells  
in each plane (each cell is two paralleled wires).  The cells are 14 mm in 
width with a $\pm$10 mm separation from the high voltage planes.  The planes 
have a total detection area of 1.25~$m^2$.    
Four high voltage supplies each furnish 
a quarter of the high 
voltage and run the 256 proportional wire planes with a 
total of 20 milliamps of current. 
Isolation resistors are used for each hut and each PWC within the hut 
so that if a wire breaks in a plane and shorts that chamber out, only 
that hut becomes dead and the rest of the array operates normally.  
If there are two shorted PWCs in a single quarter at the same time, 
then enough current is drawn from the power supply for that quarter that 
it will trip off; this will cause one quarter of the array to become 
dead, again allowing the remaining 3/4 of the array to operate normally.
Because of the uniformity in characteristics of the planes, each quarter 
of the experiment is run with all of the PWC planes at the same high voltage; 
thus no voltage divider boxes are used. 

The precision of the dimensional tolerences which were achieved by mass 
production construction techniques 
allows stable PWC operation over long periods of time 
with minimal attention.  
The high voltage, AC power, gas, clock and data co-ax cables all run 
underground.  Minimal problems have been encountered with some animals 
needing to be trapped and transported to a distant place to keep 
them from eating the outer covering of the underground 
co-ax (you wouldn't think the co-ax to be 
either that accessible or that tasty).  

Each hut enclosure of a detector station has a dehumidifier and a heater.  
The heater keeps the temperature above  
10$^\circ$ C; below this temperature, differential contraction of the 
signal wires and the PWC glass can break the 
wires.  Dehumidifiers are necessary because the top and bottom plates 
of a PWCs are a styrofoam-aluminum composite; when the relative 
humidity rises above $\sim$50\%, water vapor is absorbed into the styrofoam 
causing partial conductivity and drawing  
added high voltage current.  

\section{Summary} 
GRAND was economical to construct, requires little manpower 
to operate, and the data is quite easy to analyze.  Precision alignment of 
the PWCs was relatively easy and is absolutely maintained 
with no further work or adjustments.  
The PWC detectors are reliable and require little 
maintainence.   
The mean-time-to-failure for these proportional plane detectors is 
 $\sim$1000 years; further improvement could be made easily.  
The average angular precision for a single muon track is 0.75$^\circ$ 
(projected; 3$^\circ$ for its primary) 
and for a shower primary is $\sim0.25^\circ$ (depending on 
the number of the shower tracks).  
For each track, in addition to its angular measurement, 
it is identified as muon (or electron); the muon tracks are better 
correlated with the primary. 

Many references associated with GRAND are contained in \citet{Poirier99}.  
A more complete and up-to-date list is available on \citet{GRANDweb}.  
The numerical references to papers  
appearing on the Los Alamos Web Cite are listed in \citet{xxx}. 

\begin{acknowledgements}
Thanks are extended to those participating in GRAND's construction, 
operation, and analyses.  
GRAND is funded through the University of Notre Dame and private 
donations.
\end{acknowledgements}

%
 \begin{figure}[t]
 \vspace*{-12.0mm} 
 \includegraphics[width=8.3cm]{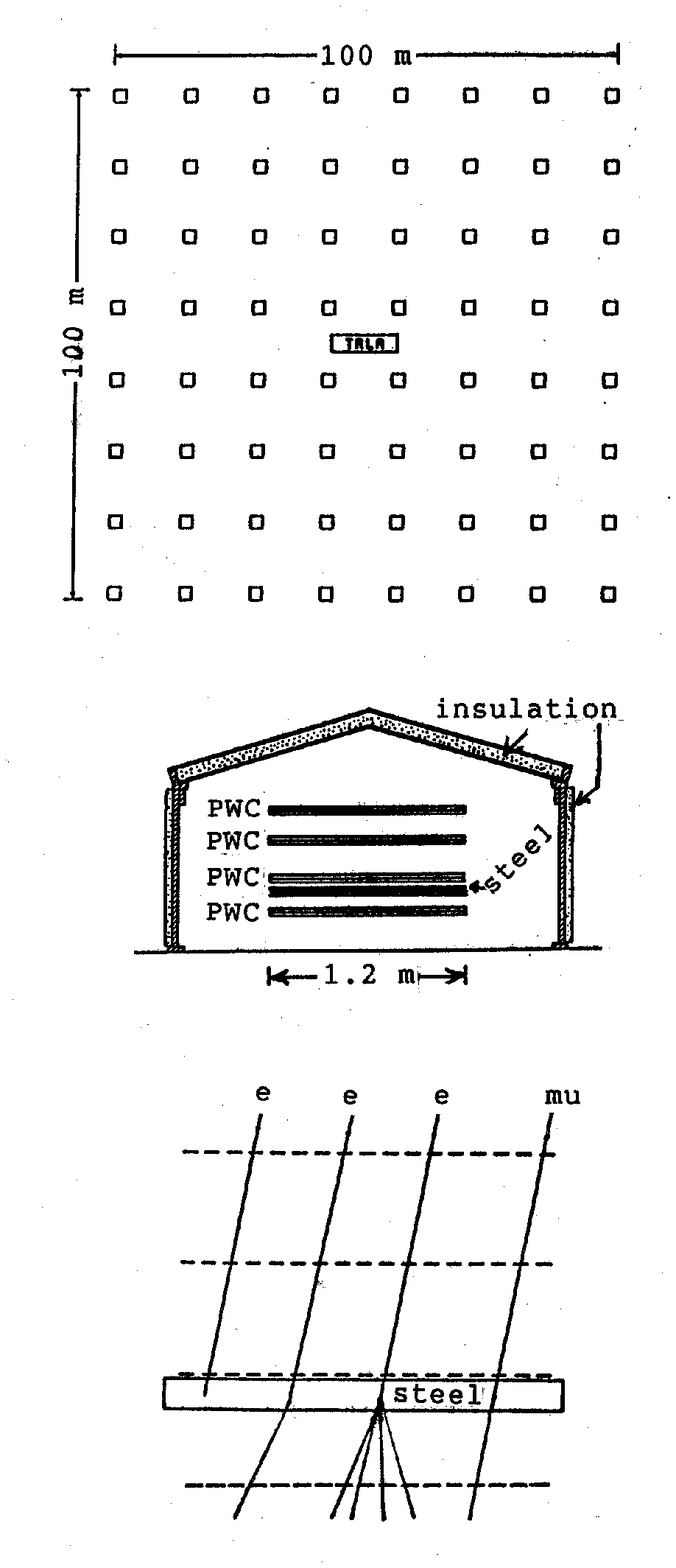} 
 \caption{Top: 100 m x 100 m field of 64 stations and the central 
 electronics trailer.  Mid: Vertical cross section of a detector 
 hut; "PWC" denotes two orthogonal proportional planes.  Bottom: 
 Schematic of a muon track and three possibilities for electron tracks.}
 \end{figure}
 
\end{document}